\documentstyle[11pt,IAUS215,twoside,epsf]{article}

\def\edcomment#1{\iffalse\marginpar{\raggedright\sl#1\/}\else\relax\fi}
\reversemarginpar

\begin{document}
\vspace*{1cm}
\title{Looking Deep Within an A-type Star: Core Convection Under the Influence of Rotation}
 \author{Allan Sacha Brun\altaffilmark{1}, Matthew Browning and Juri Toomre}
\affil{JILA, University of Colorado, Boulder, CO 80309-0440, USA}
\altaffiltext{1}{\sl New permanent address: DSM/DAPNIA/SAp, CE Saclay, 91191 Gif sur Yvette, France.}

\begin{abstract}
The advent of massively parallel supercomputing has begun to permit
explicit 3--D simulations of turbulent convection occurring within the
cores of early-type main sequence stars.  Such studies should
complement the stellar structure and evolution efforts that have so far
largely employed 1--D nonlocal mixing length descriptions for the
transport, mixing and overshooting achieved by core convection.  We
have turned to A-type stars as representative of many of the dynamical
challenges raised by core convection within rotating stars.  The
differential rotation and meridional circulations achieved deep within
the star by the convection, the likelihood of sustained magnetic dynamo
action there, and the bringing of fresh fuel into the core by
overshooting motions, thereby influencing main sequence lifetimes, all
constitute interesting dynamical questions that require detailed
modelling of global-scale convection. Using our anelastic spherical harmonic 
(ASH) code tested on the solar differential rotation problem, 
we have conducted a series of 3--D spherical domain simulations 
that deal with a simplified description of the central regions 
of rotating A-type stars, i.e a convectively unstable core is surrounded 
by a stable radiative envelope.  A sequence of 3--D simulations are 
used to assess the properties of the convection (its global patterns, 
differential rotation, meridional circulations, extent and latitudinal 
variation of the overshooting) as transitions are made between laminar 
and turbulent states by changing the effective diffusivities, rotation rates, 
and subadiabaticity of the radiative exterior.  We report on the properties 
deduced from these models for both the extent of penetration and the 
profile of rotation sustained by the convection.
\end{abstract}

\section{A Model for Core Convection}

The anelastic spherical harmonics (ASH) code solves the 3--D MHD 
anelastic equations of motion in a rotating spherical shell geometry 
using a pseudo-spectral/semi-implicit method (Miesch et al. 2000, 
Brun \& Toomre 2002, 2003). The model is a simplified description 
of an A-type star core convection zone: values for a two solar mass star are 
taken for the heat flux, nuclear energy generation (based on the CNO cycles), 
radius ($R_*\sim 2R_{\odot}$, where $R_{\odot}$ is the solar radius), 
and a perfect gas is assumed. The reference rotation rate $\Omega_0$ is 
equal to or twice the solar rotational rate (i.e. the period $P\sim$ 28 or 14 days). 
The computational domain extends from 0.02 to 0.3 $R_*$, thereby 
concentrating on the bulk of the convective core with penetration 
into the radiative envelope. The steep entropy gradient between the 
convection and radiation boundary ($r\sim 0.15 R_*$) has been softened. 
The typical density difference across the shell in radius is about 30.

\section{Topology of Evolving Convection}

Figure 1 displays two views of a 3--D rendering of the radial component of 
the velocity field. Such a representation clearly exhibits the non-spherical shape
(i.e. prolate) of the convective core. Outward motions overshoot further near the polar
regions than near the equator (cf. \S 4).

\begin{figure}[!ht]
\setlength{\unitlength}{1.0cm}
\begin{picture}(5,5.8)
\includegraphics{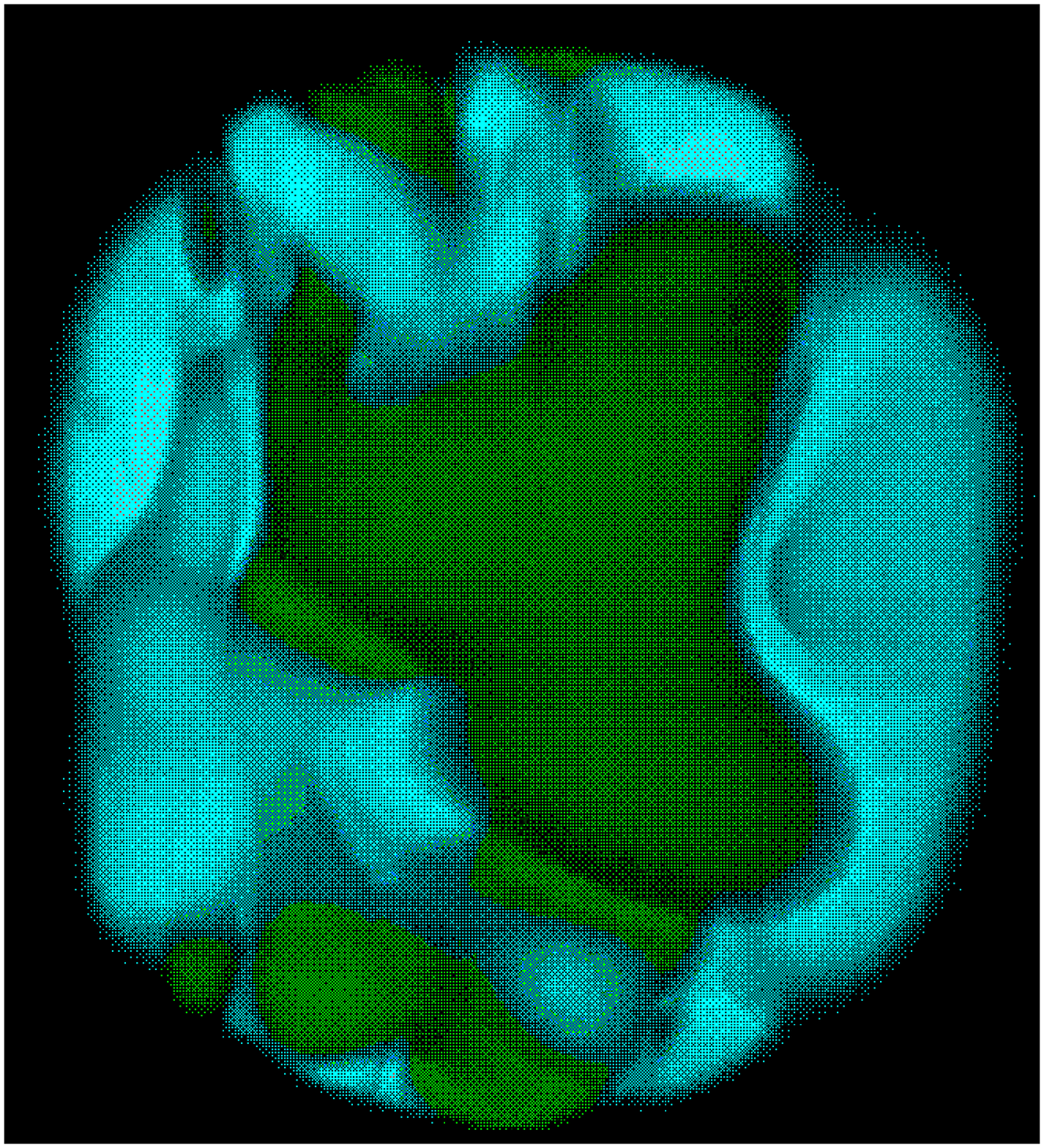}
\includegraphics{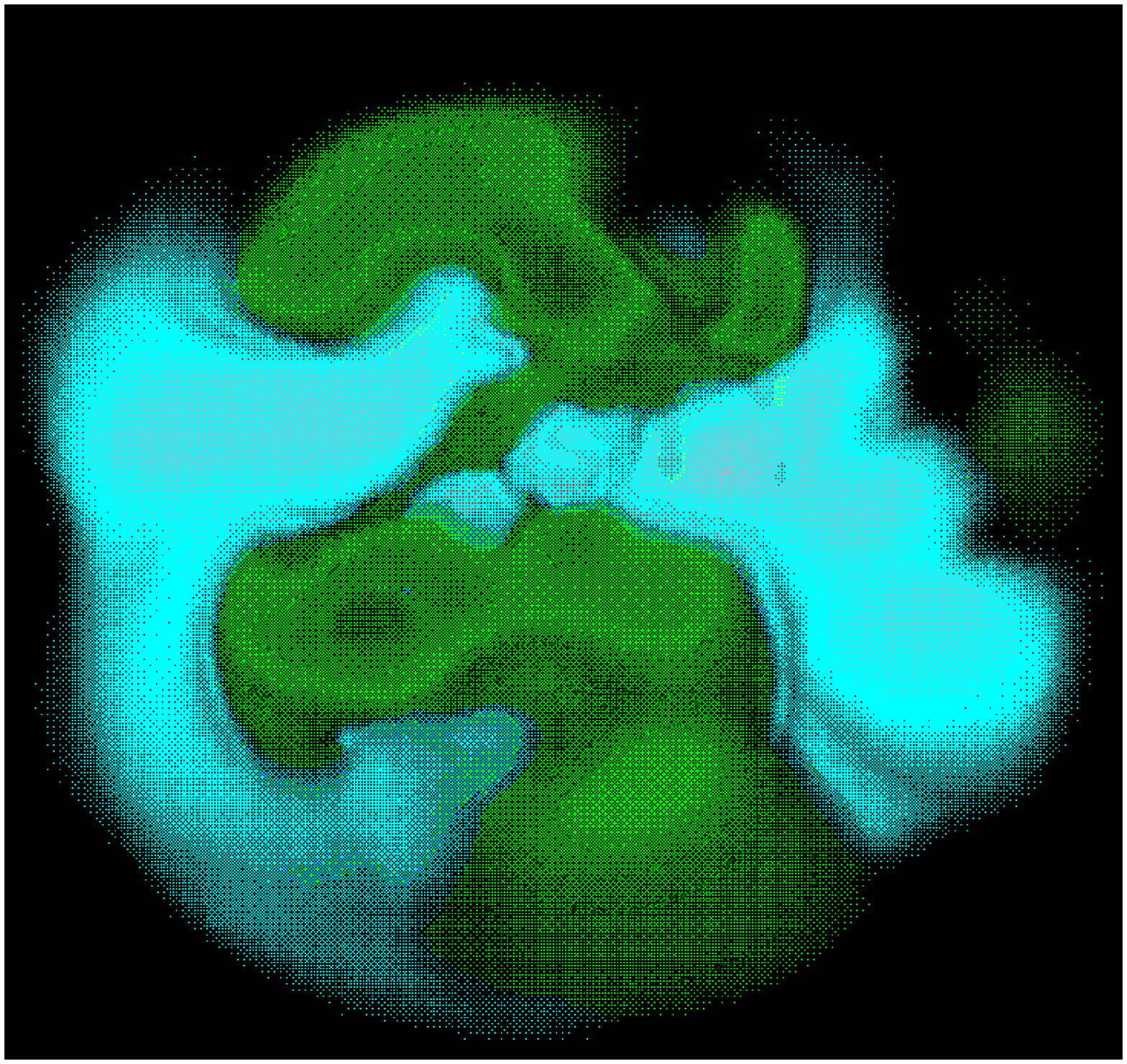}
\end{picture}
\caption[]{\label{fig1} Snapshot of the radial component of the 
velocity field for case $A1$ appearing here as a 3--D prolate shape 
seen either sideway as a whole (left) or through an equatorial cut (right). Upflows
are represented with dark tones.}
\end{figure}

Another interesting feature is the relatively large physical scale of the convective patterns. 
Effectively between our most laminar case and our most complex one, achieved
by reducing the viscous and thermal diffusivities by more than a factor of 10,
no small scales features have clearly emerged. In the latter the flow is more intricate 
and time dependent but the convective scales remain roughly the same. In addition, no obvious
asymmetries between upflows and downflows are found. Both these results are at odds 
with our solar convection simulations (Brun \& Toomre 2002) where strong downward vortical 
structures with a small filling factor (i.e. plumes) appear as the level of turbulence 
increases. One reason could be that in the convective core itself the radial density 
contrast is only of order 3 and that our runs are still too laminar. Nevertheless it 
seems that in a convective core large-scale motions are likely to be the dominant players. 
In Figure 1 (right) showing the equatorial region and southern hemisphere,
we note that the downflows and upflows are connected via the central region. As far as we can 
tell no spurious effects have been generated by the omission of the innermost core. 

\begin{figure}[!ht]
\setlength{\unitlength}{1.0cm}
\begin{picture}(5,6.5)
\includegraphics{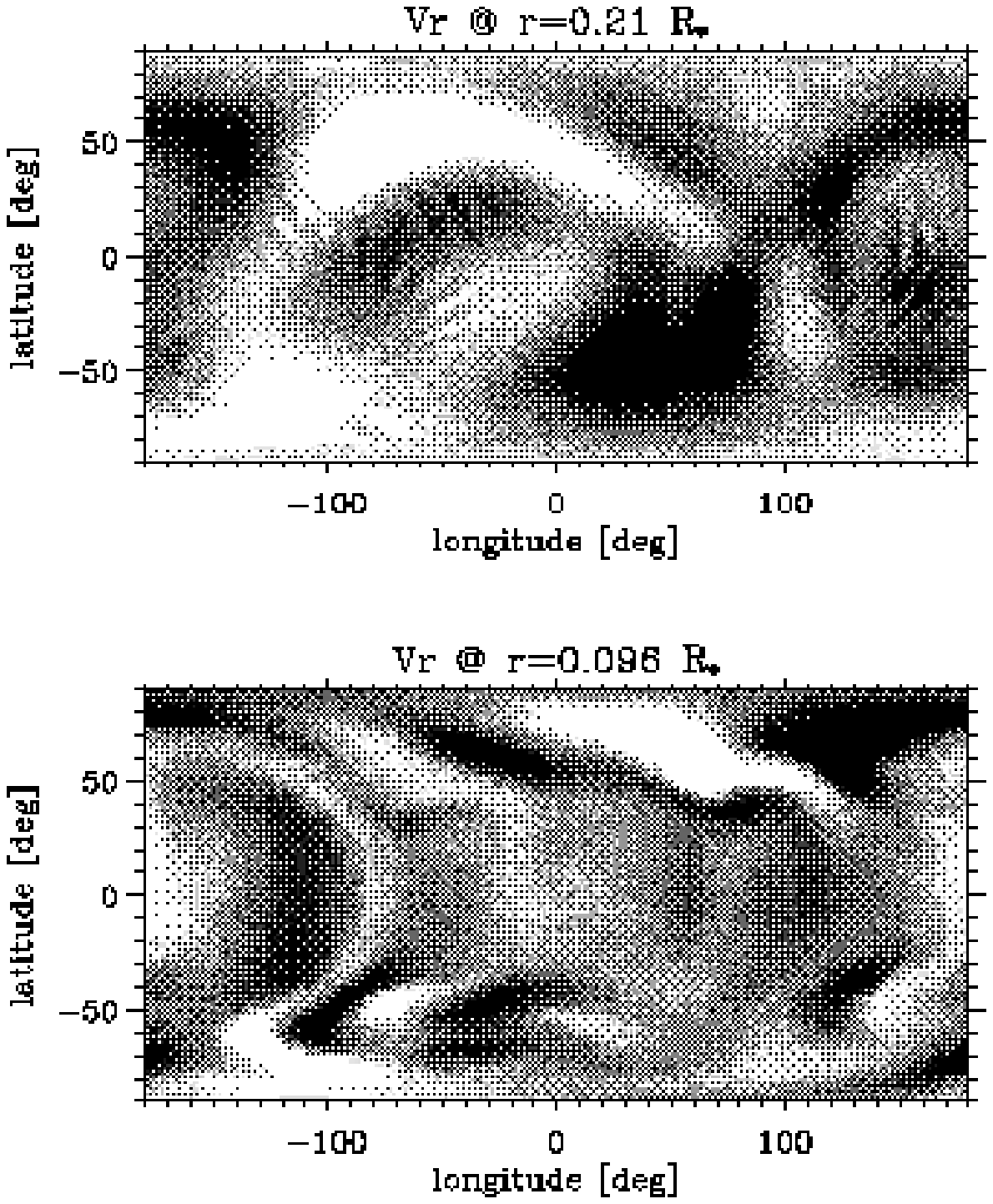} 
\includegraphics{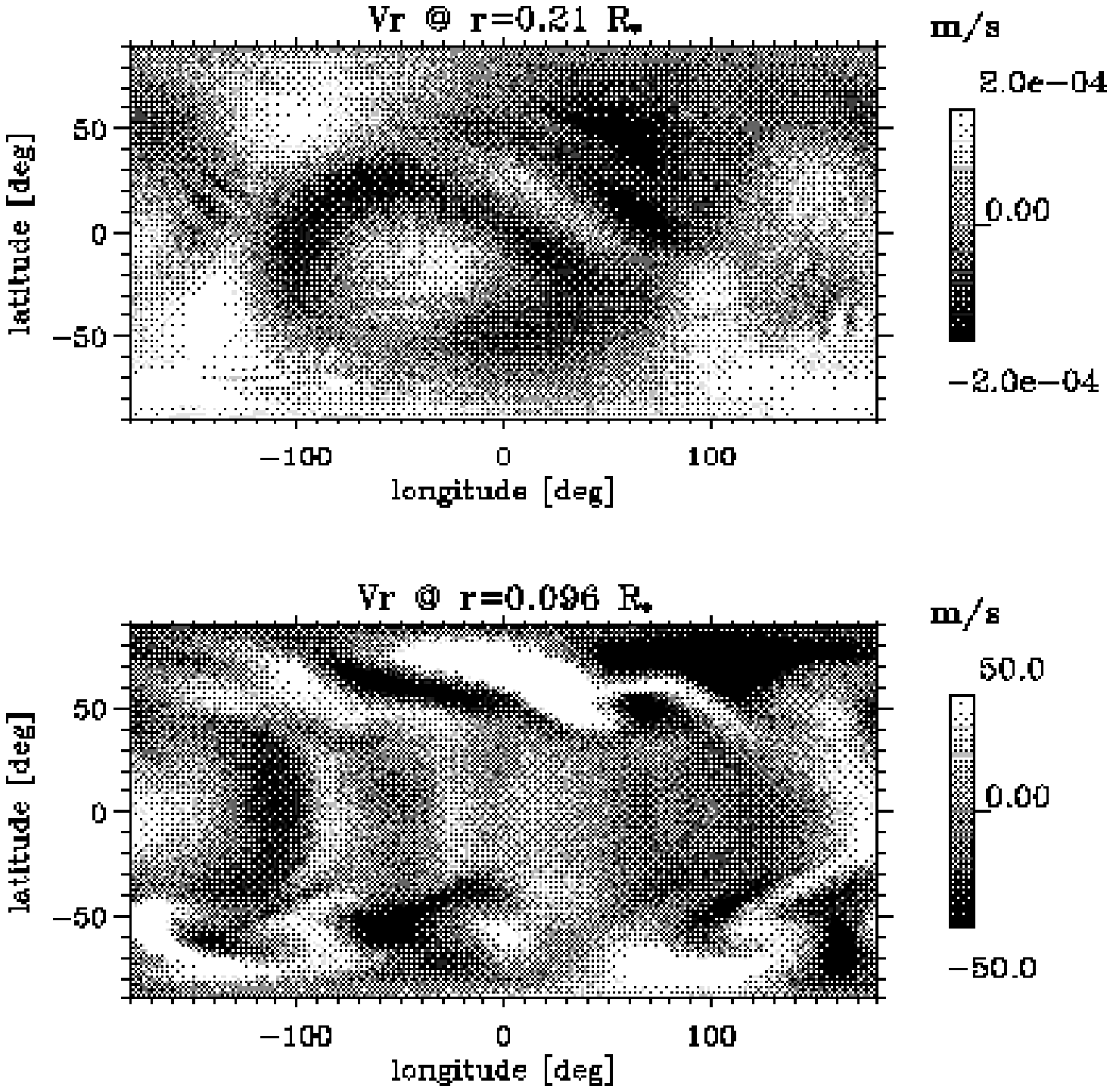} 
\end{picture}
\caption[]{\label{fig2} Two temporal samplings of the radial component of the 
velocity field for case $A1$ at mid depth in the convective core (bottom panels)
and in the radiative envelope (top). The time interval between the left and right 
set of panels is 5 days. In the upper panels the ripple patterns reveal that 
gravity waves are present. Darker tones represent downflows.}
\end{figure}

Figure 2 displays a snapshot of the radial component of the velocity field as 
a longitude and latitude map taken at mid depth in the convective (bottom) 
and radiative (top) regions. The rms radial velocities are of order 30 {\rm m/s} in the
convective core but much less in the radiative envelope. The left and right 
panels are 5 days apart and give a sense of the time evolution of the 
convective patterns. Most of the pattern deformation is due to the differential 
rotation associated with the convection.
In the upper panels located in the radiative envelope, ripples are evident. These are
gravity-inertial waves excited by the upward convective motions punching into the 
stable radiative envelope. In the case displayed, the Brunt-Va\"{\i}ss\"ala frequency 
$N/2\pi$ is roughly 30 $\mu Hz$, indicating that these waves are more gravity waves 
in nature than Rossby waves (i.e. $N/2\Omega_0>$1). Their kinetic energy spectra
peak around a wavenumber of about 30. We now turn to the properties of the
mean flows achieved in our simulations. 

\section{Rotation Profile and Circulation Achieved by the Convection}

Figure 3 display a typical averaged rotation profile as found in our core
convection simulations accompanied by latitudinal cuts of $\Omega$ and the
corresponding meridional circulation. The most striking property is a column
of slow rotation deep in the convective core leading to a significant angular
velocity gradient both in radius and latitude. At the interface between the convective
core and the radiative envelope, $\Delta \Omega/\Omega_0 \sim 40\%$ with the polar regions
being slow and the equator fast relative to the reference rotating frame.
The meridional circulation exhibits a multi-cellular structure and extends slightly
beyond the convective core; it can thus possibly bring fresh fuel inward. 
Circulations are also present in the radiative envelope but are many orders of 
magnitude weaker. These flows are time dependent, and with the Reynolds
stresses play an important role in redistributing the angular momentum. 
With our choice of stress-free top and bottom boundary conditions, the
total angular momentum is conserved in the simulations. A careful 
analysis of the angular momentum balance demonstrates that the prograde 
equatorial region is achieved due to an outward and 
equatorward transport of angular momentum by the Reynolds stresses 
that opposes the viscous and meridional circulation transports.
We also found that a positive viscous transport of angular momentum connects 
the two zones and as a consequence seems to speed up slightly the radiative envelope.
Due to the geometry of the problem, the further a fluid element is from the rotational axis,
the bigger is its angular momentum. It is a likely that the slow column at
the center may be a direct consequence of such angular momentum conservation.  
We have also found that increasing the level of complexity of the flow increases the
gradients of angular velocity. The case running with
twice the solar rotation rate also exhibits similar rotation profiles (i.e. a deep
slowly rotating column). Since most A-type stars are rotating significantly faster
than the solar rate we expect the convective motions to be even more constrained 
by rotation and thus even more aligned with the rotation axis than our cases
(Taylor columns). The consequence of such a rotation profile is that it is
likely to generate a dynamo by stretching the poloidal magnetic field into 
toroidal structures (i.e. $\omega$-effect). We refer to Browning et al. (2003) for a 
preliminary study of such effect.

\begin{figure}[!ht]
\setlength{\unitlength}{1.0cm}
\begin{picture}(5,6.2)
\includegraphics{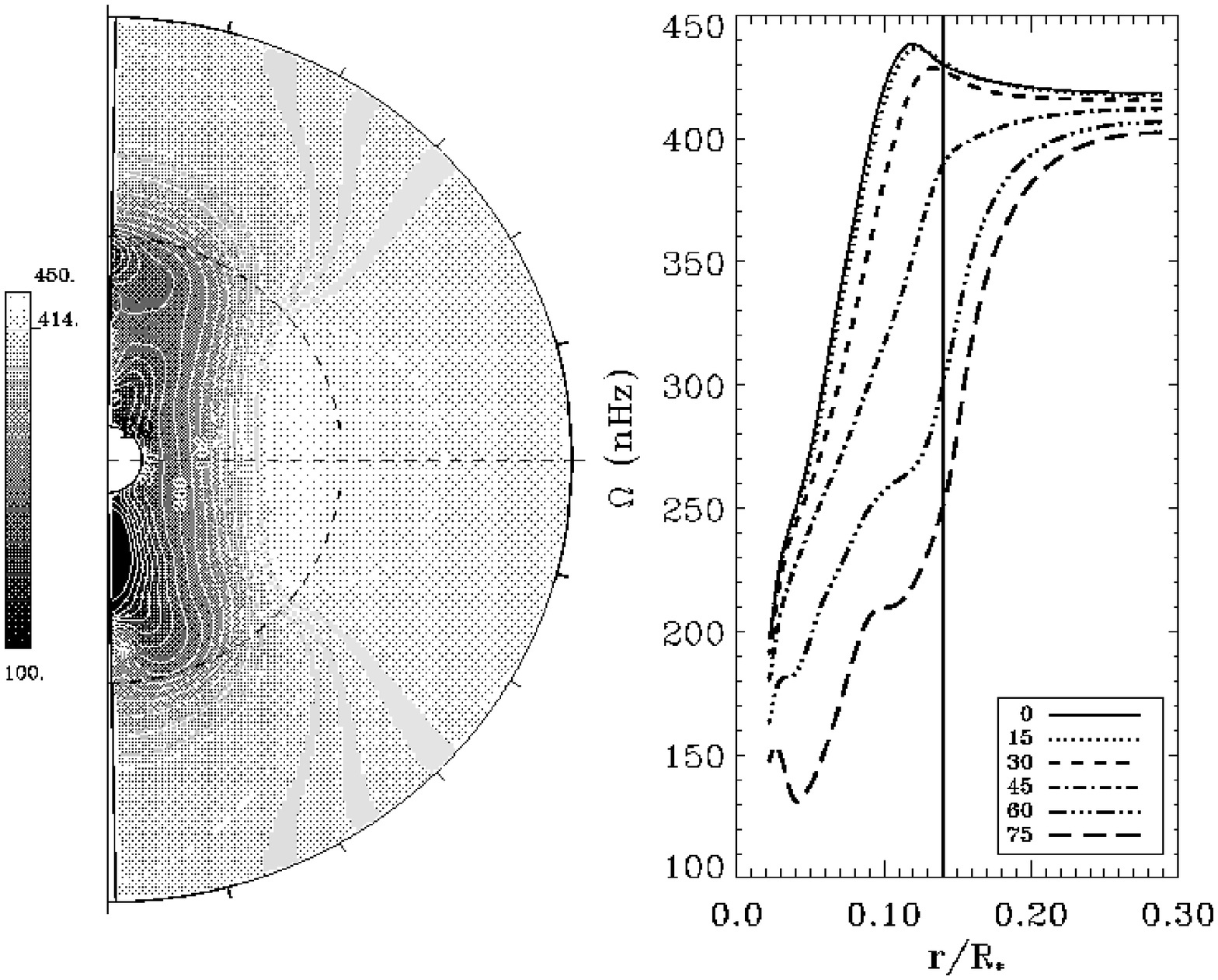}
\includegraphics{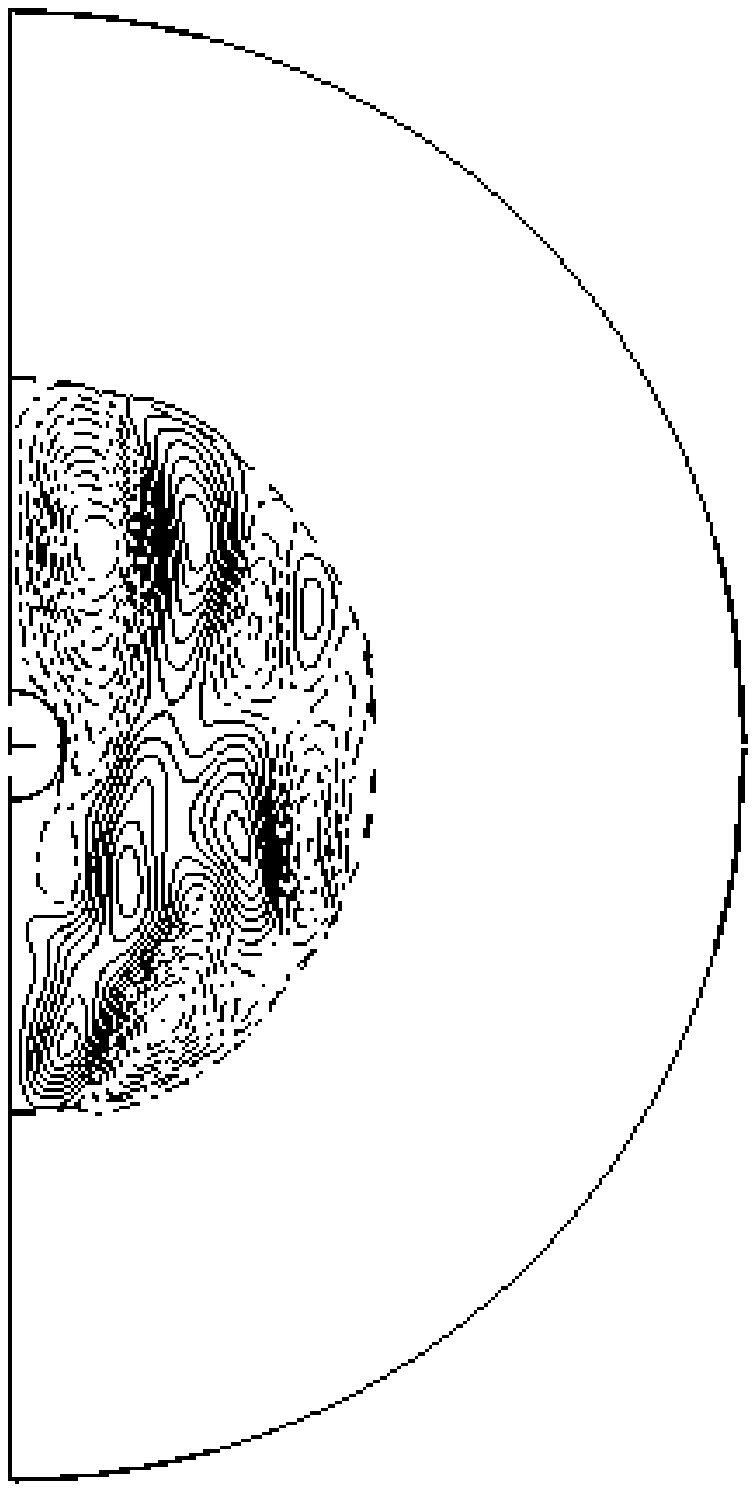}
\end{picture}
\caption[]{\label{fig3} Temporal and longitudinal averages in case A1 of the 
angular velocity profile (left panel) and of the meridional circulation (right)
formed over an interval of 85 days. This case exhibits a prograde 
equatorial rotation as well as a central region of particularly 
slow rotation. The circular dashed line delimits the convective core. 
A strong contrast $\Delta \Omega$ from equator to pole and in radius is clearly seen in the
middle panel showing latitudinal cuts of $\Omega$ as a function of radius.
The vertical solid reference line is positioned at $\nabla=\nabla_{ad}$.
The right panel shows the streamfunction in the meridional plane (with solid contours
representing counterclockwise circulation). A multi-cellular structure in radius 
and latitude of the meridional circulation is evident. Typical velocities are of 
the order of 30 {\rm m/s}. In the radiative envelope the circulations are many 
orders of magnitude smaller.}
\end{figure}

\section{Latitudinal Variation of Penetration into Radiative Envelope}

Figure 4 displays the temporal and longitudinal averages of the radial 
enthalpy (or heat) flux in the meridional plane for case $A1$ as a contour plot,
along with latitudinal cuts near the overshooting region.
This radial enthalpy flux comes about via correlations between the radial
component of the velocity field and the temperature fluctuations. 
Due to the large heat capacity of the stellar plasma
in this central region, small temperature fluctuations (of order 1 K, i.e. 
one part in 10 million) and slow motions ($\sim$ 30 {\rm m/s}) are enough
to carry outward the energy generated by the thermonuclear reactions (CNO cycles). 
Indeed around the middle of the convective core, the convective luminosity 
peaks at between 30$\%$ and 50$\%$ of the star's surface luminosity $L_*\sim 19L_{\odot}$, 
where $L_{\odot}$ is the solar luminosity; this corresponds to roughly 90$\%$ 
of the local luminosity $L(r)$, the remaining being transported by radiation.
Turning to the contour plot of Figure 4, we note that the radial enthalpy flux
is non uniform, centered in the middle part of the convective core with extrema
near the mid latitudes. This non uniform heating of the base of
the stable radiative envelope is likely to generate circulations in it.
We have indeed found small circulations in the 
radiative zone that could lead to mixing and modify the 
chemical composition on shorter time scales than for example gravitational 
settling. Another direct consequence is that we expect the overshooting to
also be nonuniform.

\begin{figure}[!ht]
\setlength{\unitlength}{1.0cm}
\begin{picture}(5,6.2)
\includegraphics{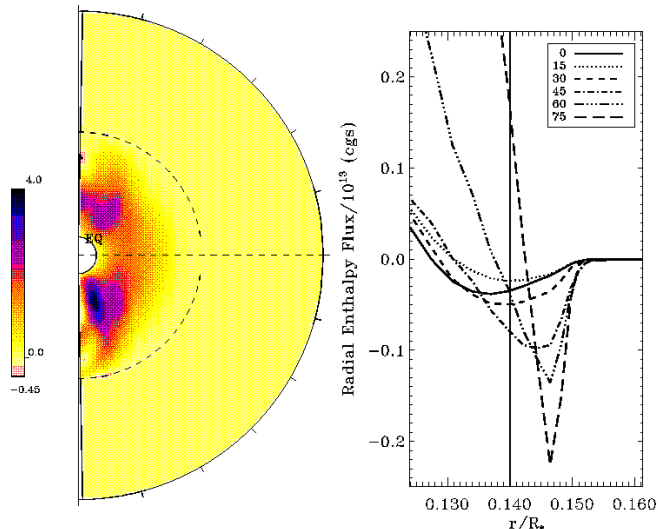}
\end{picture}
\caption[]{\label{fig4} Temporal and longitudinal averages in case $A1$ of the 
radial enthalpy flux formed over the same interval of 85 days. 
This case exhibits a nonuniform radial enthalpy flux that peaks at 
mid latitudes. The circular dashed line delimits the convective core.
Overshooting in the stable envelope is more prominent near the
poles, as in the right panel showing latitudinal cuts as a 
function of radius. The vertical reference line is positioned at $\nabla=\nabla_{ad}$.}
\end{figure}

In Figure 4 (right), we have zoomed in on the overshooting region and
plotted latitudinal cuts of the radial enthalpy flux. We believe this quantity 
to be a reasonable variable with which to assess the overshooting properties of our 
convective core. Since our ASH code deals with a single fluid, the influence 
of $\mu$-gradients on our estimates of the overshooting extent $d_{ov}$ are 
not taken into account (Canuto 1999). We first note that the overshooting
varies with latitude. Overshooting increases monotonically at a latitude of 15$\deg$
to a maximum at the poles.
This property explains why the convective core is prolate in shape.
We believe that the Coriolis force is in part responsible for
this property by deflecting the overshooting motions less 
at higher latitudes since they get closer to the rotation axis.
We have also found as expected that the stiffer the radiative envelope is, the harder
it is for the outward convective motions to overshoot. Decreasing the viscosity
of the flow leads to a smaller overshooting region. Since the actual size of the
convective patterns has not changed much between the laminar and more complex cases, 
the more complex flows are more intermitent and imprint less effectively 
upon the stable zone. We have not yet performed simulations at a sufficient
number of rotation rates to be able to deduce a reliable trend on the 
influence of the rotation rate on $d_{ov}$. We are in 
the process of computing faster rotating A-type stars. 
On average we find that for our stiffest and more complex case, the overshooting is:

\begin{equation}
d_{ov} \leq 0.012 \pm 0.003 \ R_* \mbox{ or } d_{ov} \sim 1.8\times 10^9 \mbox{\rm cm.}
\end{equation}

We consider this value as an upper limit because our simulated flows are not turbulent 
nor the stable region as stiff as in a real A-type star. We believe that 
overshooting in a convective core is certainly smaller than what is generally 
assumed in 1--D stellar evolution models of massive stars in order to obtain 
a good match between theoretical and observational isochrones (Girardi et al. 2000).

These preliminary 3--D hydrodynamical models of an A-type star convective core 
have revealed some interesting properties:
1) the convective core is differentially rotating and exhibits strong gradient in angular velocity,
2) the overshooting is nonuniform, seeming to be more prominent near the poles and 
certainly does not extend as far as generally considered in 1--D stellar models, and 
3) gravity waves and slow circulations are found in the radiative envelope that could 
mix chemical species. 

We thank N. Brummell, M. Miesch and J.-P. Zahn for most useful discussions.  
This work was partly supported by NASA through SEC Theory Program grant NAG5-8133
and by NSF through grant ATM-9731676. The simulations with ASH were carried 
out with NSF PACI support of supercomputer centers. 
The analysis of the data sets was carried out in our Laboratory of Computational
Dynamics within JILA.

\end{document}